\documentclass[reqno,12pt]{amsart}

\usepackage{amsthm,amsmath,amssymb,amsfonts}
\usepackage{fullpage}

\usepackage{graphicx}
\usepackage{float}
\usepackage{subfig}

\begin{document}

\newtheorem{thm}{Theorem}[section]
\newtheorem{cor}{Corollary}[section]
\newtheorem{lem}{Lemma}[section]
\newtheorem{prop}{Proposition}[section]
\newtheorem{rem}{Remark}[section]
\newtheorem{defn}{Definition}[section]

\def\Ref#1{Ref.~\cite{#1}}
\def\Figref#1{Fig.~\ref{#1}}

\def\const{\text{const.}}
\def\Rnum{{\mathbb R}}
\def\sgn{{\rm sgn}}
\def\bigint#1{{\displaystyle{\int_{#1}}}}
\def\i{{\rm i}}

\def\sech{{\rm sech}}
\def\cn{{\rm cn}}
\def\arccn{{\rm arccn}}
\def\sn{{\rm sn}}
\def\arcsn{{\rm arcsn}}
\def\dn{{\rm dn}}
\def\K{{\rm K}}

\def\thrdprime{{\prime\prime\prime}}
\def\frthprime{{\prime\prime\prime\prime}}

\def\s{{\sigma}}
\def\c{{\rm c}}

\def\KP{{\it KP}}
\def\loc{\text{loc}}

\allowdisplaybreaks[3]

\title{Weak compactons of nonlinearly dispersive\\ K\lowercase{d}V and KP equations}

\author{
S.C. Anco${}^1$
\lowercase{\scshape{and}}
M.L. Gandarias${}^2$
\\\\
${}^1$D\lowercase{\scshape{epartment}} \lowercase{\scshape{of}} M\lowercase{\scshape{athematics and}} S\lowercase{\scshape{tatistics}}\\
B\lowercase{\scshape{rock}} U\lowercase{\scshape{niversity}}\\
S\lowercase{\scshape{t.}} C\lowercase{\scshape{atharines}}, ON L2S3A1, C\lowercase{\scshape{anada}} \\\\
${}^2$D\lowercase{\scshape{epartment}} \lowercase{\scshape{of}} M\lowercase{\scshape{athematics}}\\
F\lowercase{\scshape{aculty of}} S\lowercase{\scshape{ciences}}, U\lowercase{\scshape{niversity of}} C\lowercase{\scshape{\'adiz}}\\
P\lowercase{\scshape{uerto}} R\lowercase{\scshape{eal}}, C\lowercase{\scshape{\'adiz}}, S\lowercase{\scshape{pain}}, 11510\\
}

\date{}

\begin{abstract}
A weak formulation is devised for the $K(m,n)$ equation 
which is a nonlinearly dispersive generalization of the gKdV equation 
having compacton solutions. 
With this formulation, explicit weak compacton solutions are derived, 
including ones that do not exist as classical (strong) solutions. 
Similar results are obtained for a nonlinearly dispersive generalization of the gKP equation in two dimensions, 
which possesses line compacton solutions. 
\end{abstract}

\maketitle

\emph{keywords}: compacton; weak solution; travelling wave; cut off; nonlinear dispersion; KdV equation; $K(m,n)$ equation; line compacton; KP equation

\section{Introduction}

In the study of dispersive nonlinear wave equations,
solutions of basic importance are travelling waves
\begin{equation}\label{travelwave}
u=U(\xi),
\quad
\xi=x-ct
\end{equation}
where $c$ is the wave speed.
Stable travelling waves for which the profile $U$ goes to zero for large $|\xi|$
describe solitary waves.
Physical and mathematical features of such solutions have been extensively studied
for a large number of dispersive nonlinear wave equations. 

Another kind of travelling wave which has attracted much interest are
compactons that describe waves with compact spatial support
--- see the review article \cite{RosZil2018} and references therein. 
For a given wave equation,
a compacton is a strong (classical) solution \eqref{travelwave} 
whose profile is, without loss of generality via translation, 
supported on $-L\leq\xi\leq L$ and vanishing for $|\xi|>L$, 
with $L>0$. 
Such solutions can be constructed by starting with a travelling wave 
such that $U(\xi)$ and derivatives of $U(\xi)$ that appear in the equation 
each vanish at $\xi=\pm L$, 
which allows extending $U(\xi)$ to be $0$ outside of $-L\leq\xi\leq L$. 

Compactons were first found for a nonlinearly-dispersive generalization of
the Korteweg-de Vries (KdV) equation, 
known as the $K(m,n)$ equation \cite{RosHym},
\begin{equation}\label{Kmn}
u_t+a(u^m)_x+b(u^n)_{xxx}=0
\end{equation}
where $m,n>0$ are the nonlinearity powers (with $m\neq1$), 
and $a,b\neq0$ are the coefficients. 
For $n=1$, this is the gKdV equation,
which reduces to the KdV equation when $m=2$.
Nonlinear dispersion arises for $n\neq 1$. 
The simplest example of a compacton has a profile given by
a cosine function
$U = \tfrac{4c}{3}\cos^2(\tfrac{1}{4}\xi)$ with $L=2\pi$
in the case $m=n=2$ for $a=b=1$ \cite{RosHym}, 
where $U$ is $C^3$.
Another simple example is
$U = \big(\tfrac{4}{3c}\cos^2(\tfrac{\sqrt{c}}{4}\xi)\big)^2$ with $L=\pi/2$
in the case $m=\tfrac{1}{2}$, $n=1$ for $a=-b=1$,
which has linear dispersion but sub-linear convection \cite{Ros2007,PelSluKokPel}. 

However, many papers in the literature focus on compact wave profiles that are only $C^0$ or $C^1$; 
it is crucial to emphasize that, for any given dispersive nonlinear wave equation, 
such profiles are not classical solutions 
because they lack sufficient differentiability to satisfy the equation.
These formal piecewise expressions are not very meaningful
and cannot arise in the context of evolution of classical initial data. 

Nevertheless, as explained in the present work, 
non-smooth compactons with less differentiability than strong solutions
still make sense as generalized solutions. 
In particular, there is a well-known weak formulation
of nonlinear wave equations which can be adapted to obtain
compactons as weak solutions satisfying a dispersive nonlinear wave equation
in a pointwise distributional sense.
This makes precise a notion of distributional compactons
introduced originally in \Ref{Ros1994}.
Importantly, weakening the differentiability allows for additional solutions,
which are only $C^0$ and therefore do not exist as strong solutions.
Unlike the formal expressions mentioned previously,
weak compacton solutions as defined here
actually satisfy the given nonlinear wave equation in a suitable rigorous sense.

The weak formulation of compactons will be considered 
for the $K(m,n)$ equation \eqref{Kmn}
as well as an analogous nonlinearly-dispersive generalization of
the Kadomstev--Petviashvili (KP) equation
given by 
\begin{equation}\label{KPmn}
(u_t+a(u^m)_x+b(u^n)_{xxx})_x+\s u_{yy} =0
\end{equation}
with nonlinearity powers $m,n>0$ and $m\neq 1$, 
and coefficients $a,b\neq0$, $\s^2=1$.
This is called the \emph{\KP$(m,n)$ equation}.
It reduces to the KP equation when $n=1$ and $m=2$. 
The KP equation possesses line solitary waves \cite{KadPet}, 
which are 2 dimensional counterparts of KdV solitary waves. 
Correspondingly, a \emph{line compacton} is a travelling wave of the form
\begin{equation}\label{linetravelwave}
u=U(\xi),
\quad
\xi = x +\mu y - \nu t
\end{equation}
such that its profile $U$ has compact support.
The parameters $(\mu,\nu)$ determine both the direction and the speed of the wave.
Specifically, $c=\nu/\sqrt{1+\mu^2}$ is the speed,
and $\theta=\arctan(\mu)$ is the angle with respect to the positive $x$ axis.

A complete classification of compacton strong solutions for the $\KP(m,n)$ equation
has been obtained recently in \Ref{AncGan23}. 
This yielded many types of explicit compactons whose profiles are given by
powers of several different functions ---
cosine and sine, Jacobi cn and sn, and quadratics. 
The main results of the present work consist of finding:
\begin{itemize}
\item
a rigorous notion of weak compactons 
\item
additional explicit $K(m,n)$ compactons (weak and strong) with different profiles 
\item
less restrictive existence conditions for $K(m,n)$ compactons as weak solutions
\item
all $K(m,n)$ compactons have two-dimensional counterparts given by $\KP(m,n)$ line compactons 
\end{itemize}

In addition, 
a variational principle will be obtained for the travelling waves \eqref{travelwave} and \eqref{linetravelwave},
despite the lack of any variational principle for 
the $K(m,n)$ equation \eqref{Kmn} nor the $\KP(m,n)$ equation \eqref{KPmn}
when $n\neq 1$.

Throughout, equations \eqref{Kmn} and \eqref{KPmn}
will be considered respectively for $x$ on the line and $x,y$ on the plane.

\section{Travelling wave ODE and solution profiles}\label{sec:travellingwaves}

Travelling waves \eqref{travelwave} 
for the $K(m,n)$ equation \eqref{Kmn} are given by the nonlinear third-order ODE 
\begin{equation}\label{U.ode}
(-gU+a U^m + b(U^n)'')'=0
\end{equation}
where
\begin{equation}\label{Kmn.g}
g=c .
\end{equation}
A similar fourth-order ODE
\begin{equation}\label{U.ode.KPmn}
(-gU+a U^m + b(U^n)'')''=0
\end{equation}
arises for line travelling waves \eqref{linetravelwave}
for the $\KP(m,n)$ equation \eqref{KPmn},
with 
\begin{equation}\label{KPmn.g}
g=\nu-\s \mu^2 = c|\sec\theta| -\s \tan^2\theta .
\end{equation}
Note that every solution of the third-order ODE \eqref{U.ode}
will satisfy the fourth-order ODE \eqref{U.ode.KPmn}. 

As shown in \Ref{AncGan23},
it is convenient to change variables 
\begin{equation}\label{V.rel}
U = V^{1/n} .
\end{equation}
Integration of the ODE \eqref{U.ode} with respect to $\xi$
followed by use of the integrating factor $V'$
yields a first-order separable ODE for $V(\xi)$
\begin{equation}\label{reducedODE.V}
V'{}^2 = E +C V + B V^{1+1/n} -A V^{1+m/n}
\end{equation}
where $E$ and $C$ are constants of integration,
and where
\begin{equation}\label{Kmn.rels}
B = \tfrac{2ng}{(n+1)b},
\quad
A =  \tfrac{2na}{(m+n)b} .
\end{equation}
ODE \eqref{reducedODE.V} has the form of the energy equation of a nonlinear oscillator
\begin{equation}\label{oscilODE.V}
V'' + \tfrac{a}{b} V^{m/n} - \tfrac{g}{b} V^{1/n} =\tfrac{1}{2}C 
\end{equation}
which admits the variational principle 
\begin{equation}
\frac{\delta S[V]}{\delta V} =0, 
\quad
S[V] = \int_\Rnum ( V'{}^2 +A V^{1+m/n} - B V^{1+1/n}  )\,d\xi - C \int_\Rnum V\,d\xi . 
\end{equation}
This is a hidden variational structure in the sense that 
neither the $K(m,n)$ equation \eqref{Kmn} nor the $\KP(m,n)$ equation \eqref{KPmn}
have a variational principle (even after introduction of a potential) when $n\neq 1$,
and the integral terms in $S[V]$ are not related to any conservation laws of these equations.
It will not be needed for the subsequent weak formulation, 
but it will be used later for the numerical solutions in section~\ref{sec:numerical}.
Issues due to the lack of smoothness of the change of variable \eqref{V.rel} for general $n$
will be addressed by suitable conditions when they arise. 

The nonlinear oscillator ODE \eqref{oscilODE.V} 
possess a wide variety of solutions that vanish at the endpoints of
an interval $-L\leq\xi\leq L$.
Under suitable conditions on $n$ and $m$, these solutions $V(\xi)$
can be cutoff at $\xi=\pm L$ to get a compacton strong solution 
whose profile is centered at $\xi=0$ and vanishes for $|\xi|\geq L$. 
Each such solution yields a corresponding compacton strong solution $U=U_\c(\xi)$ of 
both of the ODEs \eqref{U.ode}--\eqref{Kmn.g} and \eqref{U.ode.KPmn}--\eqref{KPmn.g}
through relation \eqref{V.rel}. 
In turn, $u=U_\c(\xi)$ will be a strong compacton solution of 
$K(m,n)$ equation \eqref{Kmn} and the $\KP(m,n)$ equation \eqref{KPmn},
where $\xi$ is the respective travelling wave variable \eqref{travelwave} and \eqref{linetravelwave}.

\subsection{Compacton profiles}

Corresponding compactons of the $K(m,n)$ equation \eqref{Kmn}
and the $\KP(m,n)$ equation \eqref{KPmn} have the form 
\begin{equation}\label{cutoff.soln}
u=U_\c(\xi) = U(\xi)H(L-|\xi|)
\end{equation}
where $H(\xi)$ is the Heaviside step function.
A complete classification of compactons \eqref{cutoff.soln},
which exist as strong solutions, 
is presented in \Ref{AncGan23} for the $\KP(m,n)$ equation.
Solutions were obtained with explicit profiles $U(\xi)$ of the following types:
\begin{subequations}\label{types}
\begin{align}
\text{algebraic:}\quad&
\alpha\,(1-\beta\xi^2)^p
\label{zsq}
\\
\text{cosine:}\quad&
\alpha\cos(\beta\xi)^p
\label{cos}
\\
\text{sine:}\quad&
\alpha\sin(\beta\xi)^p
\label{sin}
\\
\text{Jacobi cn:}\quad& 
\alpha\,\cn(\beta\xi,k)^p,
\quad
\alpha\,\cn(\beta(\xi +\tfrac{1}{2}L),k)^p
\label{cn}
\\
\text{Jacobi sn:}\quad& 
\alpha\,\sn(\beta\xi,k)^p,
\quad
\alpha\,\sn(\beta(\xi +L),k)^p 
\label{sn}
\end{align}
\end{subequations}
which vanish at $\xi=\pm L$ for suitably chosen $L$, 
where $\alpha$, $\beta$ are real constants, 
$k$ is a real or imaginary constant, and $p$ is a positive power. 
Profiles \eqref{zsq}, \eqref{cos}, along with the first of the two profiles \eqref{cn} 
and the second of the two profiles \eqref{sn}, 
are symmetric in $\xi$,
whereas the other profiles are antisymmetric 
when $p$ is an odd integer or an odd fraction.

It is worth noting that the Jacobi $\cn$ and $\sn$ profiles \eqref{cn}--\eqref{sn}
can be written in several different ways by means of various identities
(see e.g. \Ref{AbrSte}):
\begin{equation}
\begin{gathered}
\cn(z,k)^2 + \sn(z,k)^2 =1,
\quad
\dn(z,k)^2 = 1-k^2 \sn(z,k)^2 = 1-k^2 + k^2\cn(z,k)^2 ;
\\
\sn(z,1/k) = k\,\sn(z/k,k),
\quad
\cn(z,1/k) = \dn(z/k,k) ;
\\
\sn(z,k\i) = \dfrac{\sn(\sqrt{1+k^2}z,k/\sqrt{1+k^2})}{\sqrt{1+k^2}\dn(\sqrt{1+k^2}z,k/\sqrt{1+k^2})},
\quad
\cn(z,k\i) = \dfrac{\cn(\sqrt{1+k^2}z,k/\sqrt{1+k^2})}{\dn(\sqrt{1+k^2}z,k/\sqrt{1+k^2})} . 
\end{gathered}
\end{equation}
Note that here, as in \Ref{AncGan23}, the Maple convention is used 
for the parameter in the Jacobi $\cn$ and $\sn$ functions, 
which is the square root of the standard modulus parameter for these functions. 

The following type of symmetric profile 
\begin{align}\label{newtypes}
\text{rational Jacobi cn:}\quad& 
\alpha\,\Big( \frac{\cn(\beta\xi,k) +\kappa}{\cn(\beta\xi,k) +\gamma} \Big)^q, 
\quad
\kappa,\gamma\neq 0
\end{align}
cannot be reduced to any of the previous profiles \eqref{cn} and \eqref{sn}. 
It will be considered in section~\ref{sec:explicitsolns}.

\section{Weak formulation for compacton solutions}\label{sec:conditions}

A standard weak formulation \cite{Eva} for nonlinear PDEs
will be adapted here for 
the $K(m,n)$ equation \eqref{Kmn} and the $\KP(m,n)$ equation \eqref{KPmn}.

In general, a weak formulation of a PDE with a dependent variable $u$ 
is an integral equation that is obtained by first assuming that $u$ is a strong solution, 
then multiplying the PDE by a test function 
and integrating by parts to leave as few as derivatives on $u$ as necessary 
so that $u$ has the lowest regularity possible. 
(Note that this does not require explicitly introducing a notion of weak derivatives.
Instead $u$ will simply be assumed to belong to a suitable function space.)
All boundary terms coming from the integration by parts 
are zero due to compact support of $\psi$. 

\subsection{Weak form of $K(m,n)$ equation}

\begin{defn}
Let $\psi(t,x)$ be a test function, namely smooth with compact support.
Then the weak formulation of the $K(m,n)$ equation \eqref{Kmn} 
is given by
\begin{equation}\label{weak.Kmn}
\iint_{\Rnum\times\Rnum} ( u\psi_t +a u^m \psi_x +b u^n\psi_{xxx} )\, dt\,dx=0
\end{equation}
with $u(t,x)\in L_\loc^\infty(\Rnum)\times L_\loc^{\max(m,n)}(\Rnum)$ 
being a weak solution 
if this integral equation holds for all test functions.
\end{defn}

Essentially, in applications,
$u(t,x)$ can be thought of as continuous almost everywhere in $t$
and $u(t,x)^{\max(m,n)}$ as locally integrable in $x$.

In case of travelling waves \eqref{travelwave}, 
the integral equation \eqref{weak.Kmn} can be simplified via two steps. 
First, change variables $(t,x)\to (\tau,\xi)$ where $\tau=t$, yielding
$\iint_{\Rnum\times\Rnum} ( U(\psi_\tau -c\psi_\xi) +a U^m \psi_\xi +b U^n\psi_{\xi\xi\xi} )\, d\tau\,d\xi=0$. 
Next, use 
$\int_{\Rnum} U \psi_{\tau}\,d\tau = -\int_{\Rnum} U_{\tau} \psi\,d\tau$
due to $U_\tau=0$. 
This gives 
\begin{equation}\label{weak.Kmn.pde}
\iint_{\Rnum\times\Rnum} \big( ({-c}U +a U^m) \psi_\xi +bU^n\psi_{\xi\xi\xi} \big)\,d\tau\,d\xi=0 , 
\end{equation}
whereby $u=U(\xi)$ is a weak travelling wave solution of the $K(m,n)$ equation \eqref{Kmn} 
if this integral equation holds for all test functions $\psi(\xi,\tau)$. 

Similarly, the travelling wave ODE \eqref{U.ode}--\eqref{Kmn.g}
has the weak formulation
\begin{equation}\label{weak.Kmn.ode}
\int_{\Rnum} ( ({-g}U +a U^m)\phi'  +b U^n \phi^\thrdprime )\, d\xi =0
\end{equation}
with $U(\xi)\in L_\loc^{\max(m,n)}(\Rnum)$ being a weak solution 
if this integral equation holds for all test functions $\phi(\xi)$.

Comparison of the integral equations \eqref{weak.Kmn.ode} and \eqref{weak.Kmn.pde}
shows that $U(\xi)$ is a weak solution of the travelling wave ODE \eqref{U.ode}--\eqref{Kmn.g} 
when (and only when) 
$u=U(\xi)$ is a weak solution of the $K(m,n)$ equation \eqref{Kmn}.

\begin{prop}
Weak compactons of the $K(m,n)$ equation are travelling waves 
$u=U(\xi)\in C_c^0(\Rnum)$ that satisfy the equivalent integral equations \eqref{weak.Kmn.pde} and \eqref{weak.Kmn.ode}. 
\end{prop}

These solutions $U(\xi)$ are continuous and compact in $\xi$.
They satisfy the weak form \eqref{weak.Kmn} of the $K(m,n)$ equation. 

\subsection{Weak form of $\KP(m,n)$ equation}

The preceding weak formulations extend straightforwardly 
to the $\KP(m,n)$ equation \eqref{KPmn}: 
$u(t,x,y)\in L_\loc^\infty(\Rnum)\times L_\loc^{\max(m,n)}(\Rnum^2)$ 
is a weak solution if the integral equation 
\begin{equation}\label{weak.KP}
\iiint_{\Rnum\times\Rnum^2} ( u \psi_{tx} +\sigma u \psi_{yy} +a u^m \psi_{xx} +bu^n\psi_{xxxx} )\, dt\,dx\,dy =0
\end{equation}
holds for all test functions $\psi(t,x,y)$. 
In case of line travelling waves \eqref{linetravelwave}, 
this integral equation simplifies to the form 
\begin{equation}\label{weak.KPmn.pde}
\iiint_{\Rnum\times\Rnum^2} ( ({-g} U +a U^m)\psi_{\xi\xi} +bU^n \psi_{\xi\xi\xi\xi} )\,d\tau\,d\zeta\,d\xi=0
\end{equation}
using variables $\tau=t$, $\zeta=y$,
where $g$ is expression \eqref{KPmn.g}.

The travelling wave ODE \eqref{U.ode.KPmn}--\eqref{KPmn.g}
has the similar weak formulation
\begin{equation}\label{weak.KPmn.ode}
\int_{\Rnum} ( ({-g}U +a U^m)\phi''  +b U^n \phi^\frthprime )\, d\xi =0
\end{equation}
with $U(\xi)\in L_\loc^{\max(m,n)}(\Rnum)$ being a weak solution 
if this integral equation holds for all test functions $\phi(\xi)$.
Thus, $U(\xi)$ is a weak solution of
the travelling wave ODE \eqref{U.ode.KPmn}--\eqref{KPmn.g}
when (and only when)
$u=U(\xi)$ is a weak solution of the $K(m,n)$ equation \eqref{Kmn}.

\begin{prop}
Weak compactons of the $\KP(m,n)$ equation are travelling waves 
$u=U(\xi)\in C_c^0(\Rnum)$ that satisfy the equivalent integral equations \eqref{weak.KPmn.pde} and \eqref{weak.KPmn.ode}. 
\end{prop}

These solution also satisfy the weak form \eqref{weak.KP} of the $\KP(m,n)$ equation.

\subsection{Derivation of weak compactons via cutoff conditions}

Conditions for a travelling wave classical solution $U(\xi)$
to yield a compacton weak solution $U_\c(\xi) = U(\xi)H(L-|\xi|)$
will now be derived from the respective integral equations \eqref{weak.Kmn.ode} and \eqref{weak.KPmn.ode}. 
The following preliminary result is needed.

\begin{lem}\label{lem}
For any function $f(\xi)$ that is at least $C^1$, and any test function $\phi(\xi)$,
\begin{equation}\label{integral.ident}
\int_{\Rnum} \phi'(\xi) f(\xi) H(L-|\xi|)\, d\xi = 
\phi(L) f(L) - \phi({-L})f({-L}) - \int_{\Rnum} \phi(\xi) f'(\xi) H(L-|\xi|)\, d\xi
\end{equation}
is an identity. 
\end{lem}

This can be proved with standard methods (see e.g.\ \Ref{GelShi})
by means of integration by parts and properties of the Dirac delta distribution 
$\delta(L-|\xi|)=H'(L-|\xi|)$.
Alternatively, the support property of $H(L-|\xi|)$ can be used to reduce the integral to the interval $[-L,L]$, followed by integration by parts. 

\begin{prop}
(i) 
Suppose that $U(\xi) \in C^3$ satisfies the ODE \eqref{U.ode}--\eqref{Kmn.g}
on $[-L, L]$ and boundary conditions
\begin{equation}\label{A1A2A3=0}
A_i|_{\xi=\pm L}=0,
\quad
i=1,2,3
\end{equation}
where 
\begin{equation}\label{A1A2A3}
A_1 = bU^n ,
\quad
A_2 =  b(U^n)',
\quad
A_3  =-g U +aU^m + b(U^n)'' . 
\end{equation}
Then $U_\c(\xi)$ defined by the cutoff expression \eqref{cutoff.soln}
is a weak solution of equation \eqref{weak.Kmn.ode}.
\\
(ii) 
For $U(\xi)\in C^4$ satisfying the ODE \eqref{U.ode.KPmn}--\eqref{KPmn.g} and boundary conditions \eqref{A1A2A3=0} as well as 
\begin{equation}\label{A4=0}
A_4|_{\xi=\pm L}=0
\end{equation}
where \begin{equation}\label{A4}
A_4 = (-g U +aU^m)' + b(U^n)^\thrdprime , 
\end{equation}
$U_\c(\xi)$ is a weak solution of equation \eqref{weak.KPmn.ode}.
\end{prop}

Proof: 
To proceed using Lemma~\ref{lem}, 
substitute expression \eqref{cutoff.soln}
into the integral equations \eqref{weak.Kmn.ode} and \eqref{weak.KPmn.ode} 
and use $H^m=H^n=H$ to get 
\begin{equation}
0= \int_{\Rnum} \big( ({-g}U_\c +a U_\c^m)\phi'  +b U_\c^n \phi^\thrdprime \big)\, d\xi
= \int_{\Rnum} \big( ({-g}U +a U^m)H\phi'  +b U^n H\phi^\thrdprime \big)\, d\xi
\end{equation}
and
\begin{equation}
0=\int_{\Rnum} \big( ({-g}U_\c +a U_\c^m)\phi''  +b U_\c^n \phi^\frthprime \big)\, d\xi 
= \int_{\Rnum} \big( ({-g}U +a U^m)H\phi''  +b U^n H \phi^\frthprime \big)\, d\xi =0 . 
\end{equation}
Next, a repeated application of the relation \eqref{integral.ident} yields
\begin{equation}\label{weak.Kmn.integral}
0= \int_{\Rnum} \big( ({-g} U +a U^m)' +b (U^n)^\thrdprime \big)H\phi \,d\xi
-\big(\phi'' A_1 -\phi' A_2 +\phi A_3\big)\big|_{\xi=\pm L} 
\end{equation}
and
\begin{equation}\label{weak.KPmn.integral}
0= \int_{\Rnum} \big( ({-g} U +a U^m)'' +b (U^n)^\frthprime \big)H\phi\, d\xi
-\big( \phi''' A_1 -\phi'' A_2 +\phi' A_3 + \phi A_4 \big)\big|_{\xi=\pm L} . 
\end{equation}

In equations \eqref{weak.Kmn.integral} and \eqref{weak.KPmn.integral},
the integral term vanishes when $U(\xi)$ satisfies
the respective travelling wave ODEs. 
This leaves the boundary terms. 
Since 
$\phi|_{\xi=\pm L}$, $\phi'|_{\xi=\pm L}$, $\phi''|_{\xi=\pm L}$, $\phi'''|_{\xi =\pm L}$
are arbitrary values, 
their coefficients must separately vanish.
Hence, the integral equations hold for all test functions $\phi$ iff 
conditions \eqref{A1A2A3=0} are satisfied, 
as well as condition \eqref{A4=0}
in the case of equation \eqref{weak.KPmn.integral}.
\qed

\begin{rem}
Existence conditions \eqref{A1A2A3=0} and \eqref{A4=0}
can be derived in an alternative way 
directly from the $K(m,n)$ and $\KP(m,n)$ equations
through substitution of the cutoff profile expression \eqref{cutoff.soln}
and use of the following steps:
firstly, simplify $H(L-|\xi|)^q=H(L-|\xi|)$ for any $q>0$; 
secondly, integrate by parts and use $H(\xi\pm L)'=\delta(\xi\pm L)$;
thirdly, collect coefficients of separate singular terms and set each coefficient to zero. 
This corresponds to the PDEs being formally satisfied in a pointwise distributional sense.
\end{rem}

A detailed investigations of these conditions \eqref{A1A2A3=0} and \eqref{A4=0}
yields the following main result.

\begin{thm}\label{thm:p.conds}
Suppose a compacton profile
$u=U_\c(\xi) = U(\xi)H(L-|\xi|)$
has the behaviour
\begin{equation}\label{U.asympt}
U(\xi) \sim U_0\, (L -|\xi|)^p = U_0\, (L\mp \xi)^p 
\text{ as } \xi \sim \pm L
\end{equation}
for some power $p\neq 0$,
at the end points of its support.\\
(i) It is a weak solution of the $K(m,n)$ equation, with $m,n>0$ and $m\neq 1$,
iff $U(\xi)\in C^3(\Rnum)$ satisfies the travelling wave ODE \eqref{U.ode}--\eqref{Kmn.g}
and $p$ obeys 
\begin{equation}\label{Kmn.p.conds}
p>2/n . 
\end{equation}
(ii) It is a weak solution of the $\KP(m,n)$ equation, with $m,n>0$ and $m\neq 1$,
iff $U(\xi)\in C^4(\Rnum)$ satisfies the travelling wave ODE \eqref{U.ode.KPmn}--\eqref{KPmn.g}
and $n$, $m$, $p$, $U_0$ obey at least one of
\begin{subequations}\label{KPmn.p.conds}
\begin{gather}
p>\max(1/m,3/n), 
\quad
g=0; 
\label{p.conds1}
\\
p>\max(1/m,3/n), 
\quad
m<1,
\quad
g\neq0; 
\label{p.conds2}
\\
p>\max(1,3/n), 
\quad
m>1,
\quad
g\neq0 ;
\label{p.conds3}
\\
1\geq p=2/(n-1), 
\quad
2m+1> n\geq3,
\quad
g\neq0,
\quad
U_0= \big(\tfrac{g}{2b}\tfrac{(n-1)^2}{n(n+1)}\big)^\frac{n-1}{(n-2)(n+1)} ;
\label{p.conds4}
\\
p=2/(n-m), 
\quad
n\geq3m,
\quad
g=0,
\quad
U_0 = \big(\tfrac{-a}{2b}\tfrac{(n-m)^2}{n(n+m)}\big)^\frac{1}{n-m} ;
\label{p.conds5}
\\
p=2/(n-m), 
\quad
m+2>n\geq3m,
\quad
g\neq0,
\quad
U_0 = \big(\tfrac{-a}{2b}\tfrac{(n-m)^2}{n(n+m)}\big)^\frac{1}{n-m} .
\label{p.conds6}
\end{gather}
\end{subequations}
In all six cases \eqref{KPmn.p.conds}, the inequality \eqref{Kmn.p.conds}
is satisfied.
\end{thm}

Proof:
From the endpoint behaviour \eqref{U.asympt} assumed for $U(\xi)$, 
the first and second conditions $A_1=A_2=0$ require that, 
near $\xi=\pm L$, both expressions 
\begin{equation}
A_1 \sim b U_0^n(L\mp\xi)^{pn} 
\quad\text{ and }\quad
A_2 \sim bpn U_0^n(L\mp\xi)^{pn-1}
\end{equation} 
must go to zero.
This implies $p>0$ and $pn>1$, since $n>0$. 
The third condition similarly requires that
\begin{equation}
A_3 \sim -g U_0^p(L\mp\xi)^{p} + a U_0^m(L\mp\xi)^{pm} + bpn(pn-1) U_0^n(L\mp\xi)^{pn-2}
\end{equation}
goes to zero.
This reduces to $(L\mp\xi)^{pn-2}\sim 0$ 
due to the previous two conditions combined with $m>0$,
which implies $pn>2$.

Thus, the three conditions \eqref{A1A2A3=0} jointly hold iff $pn>2$. 

The fourth condition \eqref{A4=0} is more complicated to analyze. 
Near the cut off, $A_4$ is given by the sum of the terms
\begin{equation}
-gp U_0^p(L\mp\xi)^{p-1},
\quad
apm U_0^m(L\mp\xi)^{pm-1},
\quad
bpn(pn-1)(pn-2) U_0^n(L\mp\xi)^{pn-3} 
\end{equation}
with $a\neq0$, $b\neq0$, and $p>0$. 
It will be convenient to consider separately the cases $g=0$ and $g\neq0$. 

Case $g=0$:
The second and third terms 
either each go to zero, 
which gives $pm>1$ and $pn>3$, 
or they mutually cancel, 
which holds when 
$pm-1=pn-3\leq 0$
and $apm U_0^m + bpn(pn-1)(pn-2) U_0^n=0$. 
The first case is equivalent to $p>\max(\tfrac{1}{m},\tfrac{3}{n})$,
while the second case gives 
$p=\tfrac{2}{n-m}$ and $n\geq 3m$,
and also $apm U_0^m + bpn(pn-1)(pn-2) U_0^n=0$. 

Case $g\neq0$:
The first term either goes to zero or has to cancel the third term, since $m\neq 1$. 
In the latter case, the middle term must go to zero, 
which gives $pm-1>0$ and $p-1=pn-3\leq 0$ 
as well as $gU_0^p = bn(pn-1)(pn-2)U_0^n$. 
The two inequalities imply $1\geq p =\tfrac{2}{n-1}>\tfrac{1}{m}$, 
and hence $n\geq 3$.
In the former case, 
$p$ must be greater than $1$, 
and the sum of the second and third terms must vanish.
This can happen in several different ways depending on whether $m>1$ or $m<1$.

If $m>1$, then $p>1$ implies $pm>1$, and so the second term goes to term. 
This leaves the third term which then must also go to zero. 
Hence $pn-3>0$, and all of these inequalities together then require 
$p>\max(1,\tfrac{3}{n})$. 

If $m<1$, then either the second term and the third term mutually cancel 
or they  separately go to zero. 
The latter case has $pm-1>0$ and $pn-3>0$. 
All of the inequalities together thereby require 
$p>\max(\tfrac{1}{m},\tfrac{3}{n})$. 
The former case gives $pm-1=pn-3\leq 0$
and also $apm U_0^m + bpn(pn-1)(pn-2) U_0^n=0$. 
This requires $\tfrac{1}{m}\geq p=\tfrac{2}{n-m}>1$ using $m<1$, 
which then implies $2+m>n\geq 3m$.

In each of the previous cases, the inequality $pn>2$ can be checked to hold. 
This completes the analysis of the conditions \eqref{A1A2A3=0} and \eqref{A4=0},
which concludes the proof. 
\qed

\subsection{Existence conditions for strong solutions}

In comparison, 
the conditions for existence of compacton strong solutions \eqref{cutoff.soln} 
for the $K(m,n)$ and $\KP(m,n)$ equations are simply that 
$U(\xi)$ satisfies the respective travelling wave ODE pointwise 
and that $U$, $U'$, $U''$, $U^\thrdprime$ vanish at $\xi=\pm L$
and, additionally, that $U^\frthprime$ vanishes at $\xi=\pm L$ in the case of the $\KP(m,n)$ equation. 

\begin{prop}\label{prop:p.conds.strong}
Suppose a compacton profile \eqref{cutoff.soln} 
is a weak solution of the $K(m,n)$ equation or the $\KP(m,n)$ equation, 
with the behaviour \eqref{U.asympt} at the end points of its support.\\
(i) It is a strong solution of the $K(m,n)$ equation, with $m,n>0$ and $m\neq 1$,
iff $p$ obeys the condition
\begin{equation}\label{Kmn.p.conds.strong}
p>3 . 
\end{equation}
(ii) It is a strong solution of the $\KP(m,n)$ equation, with $m,n>0$ and $m\neq 1$,
iff $p$ obeys the condition
\begin{equation}\label{KPmn.p.conds.strong}
p>4 .
\end{equation}
\end{prop}

In \Ref{AncGan23}, 
less stringent conditions were considered, which arise 
when the travelling wave ODE is multiplied by powers of $U$ 
to remove all negative powers. 
Such solutions need not be classical solutions of the unmodified ODE. 
The modified ODE, however, does not possess a weak formulation
similar to the integral equations \eqref{weak.Kmn.ode} and \eqref{weak.KPmn.ode}.
Specifically, there will be additional terms that contain $U'{}^2$ and $U'{}^3$
and lead to products of Dirac delta distributions.

\section{Explicit weak compactons}\label{sec:explicitsolns}

All weak compactons \eqref{cutoff.soln} 
obtained from the cut off of travelling wave solutions 
with the various types of profiles \eqref{types} 
will now be summarized for the $K(m,n)$ and $KP(m,n)$ equations. 
The steps consist of:\\
(1) substitute the specific form for the travelling wave into the ODE \eqref{U.ode};\\
(2) determine the solution parameters $\alpha$, $\beta$, $g$, $q$ by 
solving an overdetermined system of algebraic equations;\\
(3) extract the cutoff $L$ and the power $p$ in the behaviour \eqref{U.asympt} near the cutoff;\\
(4) impose the existence conditions in Theorem~\ref{thm:p.conds}.\\

\subsection{Solution profiles}

The preceding steps have been carried out in Maple 
and yield the following weak compacton solutions,
which have a symmetric profile. 
Hereafter, $\K$ denotes the complete elliptic integral (see e.g. \Ref{AbrSte}). 

\emph{Algebraic compactons}:

\begin{equation}\label{zsq1}
\begin{aligned}
& u = \big(\tfrac{g(3n+1)}{2 a (n+1)}\big)^{\frac{2}{n - 1}} \Big(1 - \tfrac{a^2(n+1)(n-1)^2}{2gbn(3n+1)^2}\xi^2\Big)^{\frac{2}{n-1}}  H(L-|\xi|),
\quad
L =\tfrac{3n+1}{(n-1)|a|} \sqrt{\tfrac{2n|gb|}{n+1}},
\\&
m = \tfrac{1}{2}(n+1),
\quad 
\sgn(g)=\sgn(a)=\sgn(b) 
\end{aligned}
\end{equation}

\begin{equation}\label{zsq2}
\begin{aligned}
&  u = \big(\tfrac{a (n+1)}{2 g}\big)^{\frac{1}{n-1}}
\Big( 1 +\tfrac{ g^2(n-1)^2}{abn(n +1)^2} \xi^2\Big)^{\frac{1}{n - 1}} H(L-|\xi|),
\quad
L=\tfrac{(n+1)\sqrt{n|ab|}}{(n-1)|g|} , 
\\&
m = 2 -n,
\quad
\sgn(g)=\sgn(a)=-\sgn(b)
\end{aligned}
\end{equation}

These two solutions first appeared in \Ref{AncGan23}.
A special case of the first solution can be found in \Ref{RosHym}.

\emph{Cosine compactons}:

\begin{equation}\label{cos1}
\begin{aligned}
& u = \big(\tfrac{2 n g}{(n+1)a}\big)^{\frac{1}{n-1}}
\cos\Big( \tfrac{n-1}{2 n}\sqrt{\tfrac{a}{b}}\xi \Big)^{\frac{2}{n-1}} H(L-|\xi|),
\quad
L= \tfrac{n}{n-1}\sqrt{\tfrac{b}{a}} \pi,
\\& 
m=n,
\quad
\sgn(g)=\sgn(a)=\sgn(b) 
\end{aligned}
\end{equation}

\begin{equation}\label{cos2}
\begin{aligned}
& u = \big(\tfrac{2 a}{g (m+1)}\big)^{\frac{1}{1 - m}} \cos\Big(\sqrt{\tfrac{-g}{4b}} (1-m) \xi\Big)^{\frac{2}{1-m}} H(L-|\xi|),
\quad
L = \tfrac{1}{1 -m}\sqrt{\tfrac{|g|}{|b|}} \pi , 
\\&
n = 1,
\quad
\sgn(g)=\sgn(a)=-\sgn(b) 
\end{aligned}
\end{equation}

These solutions were obtained respectively in \Ref{Ros1997} and \Ref{Ros2007,PelSluKokPel} for the $K(m,n)$ equation with $a=b=1$ and $a=-b=1$. 
They were re-derived in \Ref{XieYan} for the $\KP(m,n)$ equation.

\emph{Jacobi $\cn$ compactons}:

\begin{equation}\label{cn1}
\begin{aligned}
& u = \big(\tfrac{a (n +1)}{g(3 n - 1)}\big)^{\frac{1}{2 - 2n}} 
\cn\Big((1-n)\sqrt{\tfrac{-a}{nb}}\sqrt[4]{\tfrac{g}{a(3n-1)(n+1)}} \xi, \tfrac{1}{\sqrt{2}}\Big)^{\tfrac{2}{1-n}} H(L-|\xi|),
\\&
L=\tfrac{\sqrt{n|b|}}{1-n} \sqrt[4]{\tfrac{(3n-1)(n+1)}{ag}} K(\tfrac{1}{\sqrt{2}}),
\\&
m = 2 n - 1,
\quad
\sgn(g)=\sgn(a)=-\sgn(b) 
\end{aligned}
\end{equation}

\begin{equation}\label{cn2}
\begin{aligned}
& u = \big(\tfrac{g(3 n - 1)}{a (n +1)}\big)^{\frac{1}{2 n - 2}} 
\cn\Big((n-1)\sqrt{\tfrac{a}{bn}} \sqrt[4]{\tfrac{g}{a(3 n - 1)(n+1)}} \xi, \tfrac{1}{\sqrt{2}}\Big)^{\frac{2}{n - 1}} H(L-|\xi|),
\\& 
L=\tfrac{\sqrt{n|b|}}{n-1} \sqrt[4]{\tfrac{(3 n - 1)(n+1)}{ag}} \K\big(\tfrac{1}{\sqrt{2}}\big), 
\\&
m=2n-1 ,
\quad
\sgn(g)=\sgn(a)=\sgn(b)
\end{aligned}
\end{equation}

\emph{Jacobi $\sn$ compactons}:

\begin{equation}\label{sn1}
\begin{aligned}
& u = \big(\tfrac{a (n +1)}{g(3 n - 1)}\big)^{\frac{1}{2  - 2n}} 
\sn\Big((1-n)\sqrt{\tfrac{-a}{2bn}} \sqrt[4]{\tfrac{g}{a(3 n - 1)(n+1)}} (\xi+L), i\Big)^{\frac{2}{1-n}} H(L-|\xi|),
\\& 
L=\tfrac{\sqrt{2n|b|}}{1-n} \sqrt[4]{\tfrac{(3 n - 1)(n+1)}{ag}} \K\big(i\big),
\\&
m=2n-1 ,
\quad
\sgn(g)=\sgn(a)=-\sgn(b)
\end{aligned}
\end{equation}

\begin{equation}\label{sn2}
\begin{aligned}
& u = \big(\tfrac{g(3 n - 1)}{a (n +1)}\big)^{\frac{1}{2 n - 2}} 
 \sn\Big((n-1)\sqrt{\tfrac{a}{2bn}} \sqrt[4]{\tfrac{g}{a(3 n - 1)(n+1)}} (\xi+L), i\Big)^{\frac{2}{n - 1}} H(L-|\xi|),
\\& 
L=\tfrac{\sqrt{2n|b|}}{n-1}\sqrt[4]{\tfrac{(3 n - 1)(n+1)}{ag}} \K\big(i\big),
\\&
m=2n-1,
\quad
\sgn(g)=\sgn(a)=\sgn(b) 
\end{aligned}
\end{equation}

The four elliptic function solutions first appeared in \Ref{AncGan23}.
(\Ref{RosHym} mentions such solutions exist for equation $K(2,3)$
but does not state them explicitly.)

The steps deriving solutions \eqref{zsq1}--\eqref{sn2} 
can be applied to the new type of profiles \eqref{newtypes}. 
This yields the following six new compacton solutions. 
(In particular, for each solution, 
$m$ differs compared to the previous Jacobi $\cn$ and $\sn$ solutions.)

\emph{Rational Jacobi $\cn$ compactons}:

\begin{equation}\label{ratcn1}
\begin{aligned}
& u = \Big(\tfrac{(5+3\sqrt{3})g (2n -1)}{-2a(n + 1)}\Big)^{\frac{1}{3n-3}} \Bigg( 
\frac{\cn\big((n-1)\sqrt[6]{\tfrac{12 \sqrt{3} ag^2}{b^3 n^3(n+1)^2(2n-1)}}(\xi+L),\tfrac{\sqrt{2}(\sqrt{3}-1)}{4}\big) - 1}{\cn\big((n-1)\sqrt[6]{\tfrac{12\sqrt{3} ag^2}{b^3n^3(n+1)^2(2n-1)}}(\xi+L),\tfrac{\sqrt{2}(\sqrt{3}-1)}{4}\big) +\sqrt{3}+2}
\Bigg)^{\tfrac{1}{n-1}} H(L-|\xi|),
\\&
L=\tfrac{2}{n-1} \sqrt[6]{\tfrac{b^3 n^3(n+1)^2(2n-1)}{12\sqrt{3} ag^2}}
\K\big(\tfrac{\sqrt{2}(\sqrt{3}-1)}{4}\big),
\\&
m = 3 n - 2,
\quad
\sgn(a)=\sgn(b)=-\sgn(g)
\end{aligned}
\end{equation}

\begin{equation}\label{ratcn2}
\begin{aligned}
& u = \Big(\tfrac{(5+3\sqrt{3})g (2n -1)}{-2a(n + 1)}\Big)^{\frac{1}{3n-3}} \Bigg( 
\frac{\cn\big((n-1)\sqrt[6]{\tfrac{12 \sqrt{3} ag^2}{b^3 n^3(n+1)^2(2n-1)}}\xi,\tfrac{\sqrt{2}(\sqrt{3}-1)}{4}\big) + 1}{\cn\big((n-1)\sqrt[6]{\tfrac{12\sqrt{3} ag^2}{b^3n^3(n+1)^2(2n-1)}}\xi,\tfrac{\sqrt{2}(\sqrt{3}-1)}{4}\big) -\sqrt{3}-2}
\Bigg)^{\tfrac{1}{n-1}} H(L-|\xi|),
\\&
L=\tfrac{2}{n-1} \sqrt[6]{\tfrac{b^3 n^3(n+1)^2(2n-1)}{12\sqrt{3} ag^2}}
\K\big(\tfrac{\sqrt{2}(\sqrt{3}-1)}{4}\big),
\\&
m = 3 n - 2,
\quad
\sgn(a)=\sgn(b) =-\sgn(g)
\end{aligned}
\end{equation}

\begin{equation}\label{ratcn3}
\begin{aligned}
& u = \Big(\tfrac{(5+3\sqrt{3})a(n + 1)}{g (2n-1)}\Big)^{\frac{1}{3-3n}} \Bigg( 
\frac{\cn\big((1-n)\sqrt[6]{\tfrac{-12 \sqrt{3} ag^2}{b^3 n^3(n+1)^2(2n-1)}}\xi,\tfrac{\sqrt{2}(\sqrt{3}+1)}{4}\big) +\sqrt{3}-2}{\cn\big((1-n)\sqrt[6]{\tfrac{-12\sqrt{3} ag^2}{b^3n^3(n+1)^2(2n-1)}}\xi,\tfrac{\sqrt{2}(\sqrt{3}+1)}{4}\big) +1}
\Bigg)^{\tfrac{1}{1-n}} H(L-|\xi|),
\\&
L=\tfrac{1}{1-n} \sqrt[6]{\tfrac{|b|^3 n^3(n+1)^2(2n-1)}{12\sqrt{3} |a|g^2}}
\arccn\big(\sqrt{3}-2,\tfrac{\sqrt{2}(\sqrt{3}+1)}{4}\big),
\\&
m = 3 n - 2,
\quad
\sgn(a)=-\sgn(b) =\sgn(g)
\end{aligned}
\end{equation}

\begin{equation}\label{ratcn4}
\begin{aligned}
& u = \Big(\tfrac{(5+3\sqrt{3})a (n +1)}{-2g(5n - 1)}\Big)^{\frac{2}{3-3n}} \Bigg( 
\frac{\cn\big((1-n)\sqrt[6]{\tfrac{-3\sqrt{3} a^2 g}{2b^3 n^3(n+1)(5n-1)^2}}(\xi+L),\tfrac{\sqrt{2}(\sqrt{3}-1)}{4}\big) - 1}{\cn\big((1-n)\sqrt[6]{\tfrac{-3\sqrt{3} a^2 g}{2b^3n^3(n+1)(5n-1)^2}}(\xi+L),\tfrac{\sqrt{2}(\sqrt{3}-1)}{4}\big) +\sqrt{3}+2}
\Bigg)^{\tfrac{2}{1-n}} H(L-|\xi|),
\\&
L=\tfrac{2}{1-n} \sqrt[6]{\tfrac{2|b|^3 n^3(n+1)(5n-1)^2}{3\sqrt{3} a^2|g|}}
\K\big(\tfrac{\sqrt{2}(\sqrt{3}-1)}{4}\big),
\\&
m = \tfrac{1}{2}(3 n - 1),
\quad
\sgn(a)=-\sgn(b)=\sgn(g)
\end{aligned}
\end{equation}

\begin{equation}\label{ratcn5}
\begin{aligned}
& u = \Big(\tfrac{(5+3\sqrt{3})a (n +1)}{-2g(5n - 1)}\Big)^{\frac{2}{3-3n}} \Bigg( 
\frac{\cn\big((1-n)\sqrt[6]{\tfrac{-3\sqrt{3} a^2 g}{2b^3 n^3(n+1)(5n-1)^2}} \xi,\tfrac{\sqrt{2}(\sqrt{3}-1)}{4}\big) + 1}{\cn\big((1-n)\sqrt[6]{\tfrac{-3\sqrt{3} a^2 g}{2b^3n^3(n+1)(5n-1)^2}}\xi,\tfrac{\sqrt{2}(\sqrt{3}-1)}{4}\big) -\sqrt{3}-2}
\Bigg)^{\tfrac{2}{1-n}} H(L-|\xi|),
\\&
L=\tfrac{2}{1-n} \sqrt[6]{\tfrac{2|b|^3 n^3(n+1)(5n-1)^2}{3\sqrt{3} a^2|g|}}
\K\big(\tfrac{\sqrt{2}(\sqrt{3}-1)}{4}\big),
\\&
m = \tfrac{1}{2}(3 n - 1),
\quad
\sgn(a)=-\sgn(b)=\sgn(g)
\end{aligned}
\end{equation}

\begin{equation}\label{ratcn6}
\begin{aligned}
& u = \Big(\tfrac{(5+3\sqrt{3})g (5n-1)}{a(n+1)}\Big)^{\frac{2}{3n-3}} \Bigg( 
\frac{\cn\big((n-1)\sqrt[6]{\tfrac{3\sqrt{3} a^2 g}{2b^3 n^3(n+1)(5n-1)^2}}\xi,\tfrac{\sqrt{2}(\sqrt{3}+1)}{4}\big) +\sqrt{3}-2}{\cn\big((n-1)\sqrt[6]{\tfrac{3\sqrt{3} a^2 g}{2b^3 n^3(n+1)(5n-1)^2}}\xi,\tfrac{\sqrt{2}(\sqrt{3}+1)}{4}\big) +1}
\Bigg)^{\tfrac{2}{1-n}} H(L-|\xi|),
\\&
L=\tfrac{1}{n-1} \sqrt[6]{\tfrac{2|b|^3 n^3(n+1)(5n-1)^2}{3\sqrt{3} a^2|g|}}
\arccn\big(\sqrt{3}-2,\tfrac{\sqrt{2}(\sqrt{3}+1)}{4}\big),
\\&
m = \tfrac{1}{2}(3 n - 1),
\quad
\sgn(a)=\sgn(b) =\sgn(g)
\end{aligned}
\end{equation}

\begin{figure}[H]
\centering
\includegraphics[trim=2cm 17cm 10cm 2cm,clip,width=0.47\textwidth]{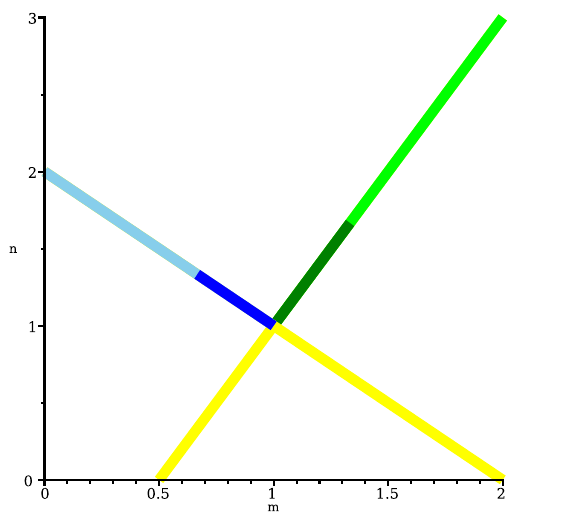}
\includegraphics[trim=2cm 17cm 10cm 2cm,clip,width=0.47\textwidth]{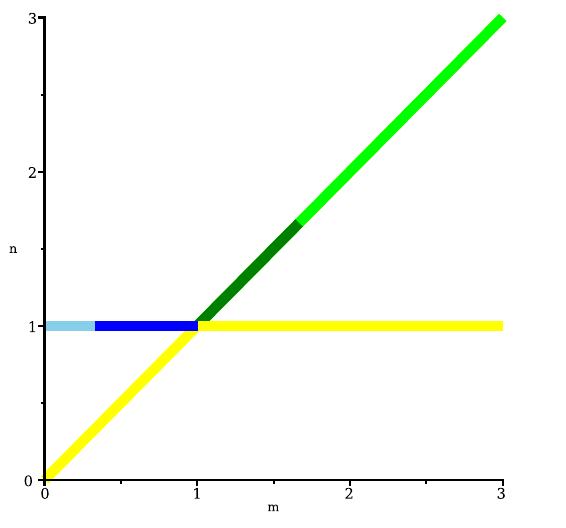}
\caption{
Parameter regions for $K(m,n)$:
(left) algebraic solutions \eqref{zsq1} [green] and \eqref{zsq2} [blue];
(right) cosine solutions \eqref{cos1} [green] and \eqref{cos2} [blue]. 
}\label{fig-alg-cos-mn}
\end{figure}

\begin{figure}[H]
\centering
\includegraphics[trim=2cm 17cm 10cm 2cm,clip,width=0.47\textwidth]{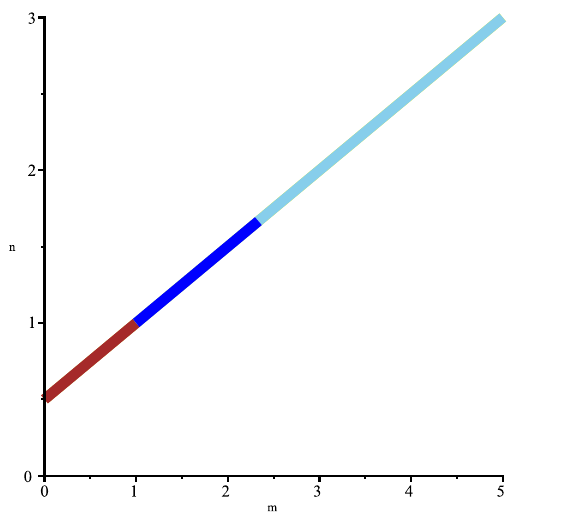}
\includegraphics[trim=2cm 17cm 10cm 2cm,clip,width=0.47\textwidth]{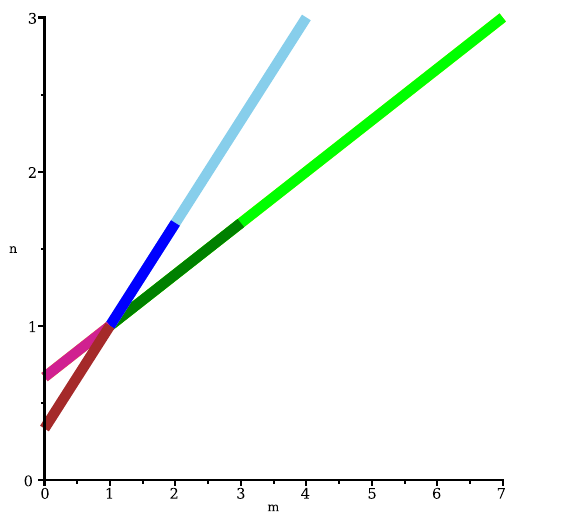}
\caption{
Parameter regions for $K(m,n)$:
(left) cn \& sn solutions \eqref{cn1} \& \eqref{sn1} [brown] and \eqref{cn2} \& \eqref{sn2} [blue];
(right) rational cn solutions \eqref{ratcn1} \& \eqref{ratcn2} [green] and \eqref{ratcn3} [pink], \eqref{ratcn4} \& \eqref{ratcn5} [brown] and \eqref{ratcn6} [blue]. 
}\label{fig-cn-ratcn-mn}
\end{figure}

\begin{figure}[H]
\centering
\includegraphics[trim=2cm 17cm 10cm 2cm,clip,width=0.47\textwidth]{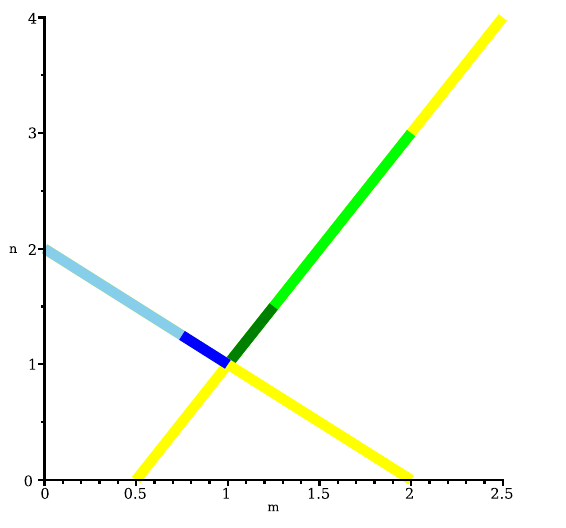}
\includegraphics[trim=2cm 17cm 10cm 2cm,clip,width=0.47\textwidth]{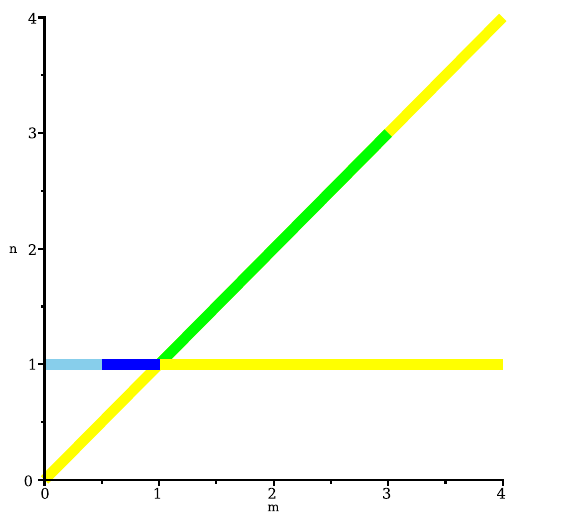}
\caption{
Parameter regions for $\KP(m,n)$: 
(left) algebraic solutions \eqref{zsq1} [green] and \eqref{zsq2} [blue];
(right) cosine solutions \eqref{cos1} [green] and \eqref{cos2} [blue]. 
}\label{fig-KP-alg-cos-mn}
\end{figure}

\begin{figure}[H]
\centering
\includegraphics[trim=2cm 17cm 10cm 2cm,clip,width=0.47\textwidth]{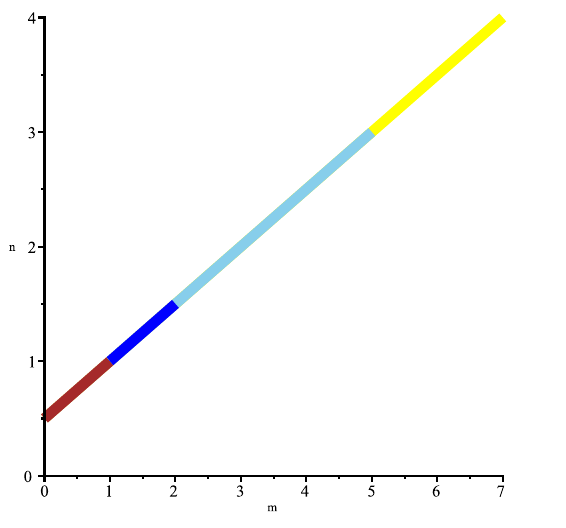}
\includegraphics[trim=2cm 17cm 10cm 2cm,clip,width=0.47\textwidth]{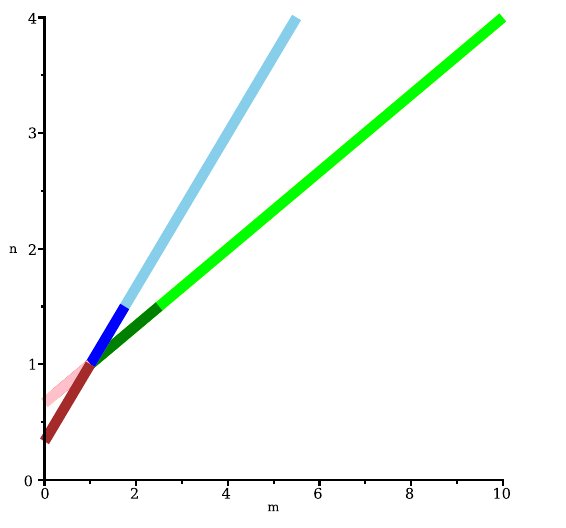}
\caption{
Parameter regions for $\KP(m,n)$: 
(left) cn \& sn solutions \eqref{cn1} \& \eqref{sn1} [brown] and \eqref{cn2} \& \eqref{sn2} [blue];
(right) rational cn solutions \eqref{ratcn1} \& \eqref{ratcn2} [green] and \eqref{ratcn3} [pink], \eqref{ratcn4} \& \eqref{ratcn5} [brown] and \eqref{ratcn6} [blue]. 
}\label{fig-KP-cn-ratcn-mn}
\end{figure}

All of these compactons \eqref{cos1}--\eqref{sn2}
exist as strong and weak solutions under suitable conditions. 
As seen in Table~\ref{table:conditions},
the weak existence conditions are less restrictive than the strong existence conditions
for the following solutions:
\eqref{zsq1}--\eqref{cos2}, \eqref{cn2}, \eqref{sn2}, \eqref{ratcn1}, \eqref{ratcn2}, \eqref{ratcn6};
for the other solutions, 
the weak conditions imply the strong conditions.
In particular, the allowed region in the parameter space $(m,n)$
for each solution is shown
in Figs.~\ref{fig-alg-cos-mn} and~\ref{fig-cn-ratcn-mn} for the $K(m,n)$ equation,
and in  Figs.~\ref{fig-KP-alg-cos-mn} and~\ref{fig-KP-cn-ratcn-mn} for the $\KP(m,n)$ equation,
where the weak and strong regions are respectively denoted by light and dark colours, while yellow denotes the line defined by the relation between $m$ and $n$.

The four classes of solutions
--- algebraic, cosine, cn \& sn, rational cn ---
have distinct values of $(m,n)$.
Within the rational cn class,
solutions \eqref{ratcn1} and \eqref{ratcn2}
exist for the same values of $(m,n)$,
and likewise for solutions \eqref{ratcn4}--\eqref{ratcn5}. 

Similar results hold for compactons that have an antisymmetric profile: 
\eqref{sin} and the second of \eqref{cn} as well as the first of \eqref{sn},
when $p$ is an odd integer or an odd fraction.

\begin{table}
\hbox{\hspace{-0.5in}
\begin{tabular}{c||c|c|c|c|c|c}
\hline
type & profile & $p$ & weak $K(m,n)$ & strong $K(m,n)$ & weak $\KP(m,n)$ & strong $\KP(m,n)$ 
\\
\hline
\hline
algebraic 
& \eqref{zsq1}
& $\tfrac{2}{n-1}$
& $1<n$
& $1<n<\tfrac{5}{3}$
& $1<n<3$
& $1<n<\tfrac{3}{2}$
\\
& \eqref{zsq2}
& $\tfrac{1}{n-1}$
& $1<n<2$
& $1<n<\tfrac{4}{3}$
& $1<n<2$
& $1<n<\tfrac{5}{4}$
\\
\hline
cosine 
& \eqref{cos1}
& $\tfrac{2}{n-1}$
& $1<n$
& $1<n<\tfrac{5}{3}$
& $1<n<3$
& $1<n<\tfrac{3}{2}$
\\
& \eqref{cos2}
& $\tfrac{2}{1-m}$
& $0<m<1$
& $\tfrac{1}{3}<m<1$
& $0<m<1$
& $\tfrac{1}{2}<m<1$
\\
\hline
cn 
& \eqref{cn1} 
& $\tfrac{2}{1-n}$
& $\tfrac{1}{2}<n<1$
& $\tfrac{1}{2}<n<1$
& $\tfrac{1}{2}<n<1$
& $\tfrac{1}{2}<n<1$
\\
& \eqref{cn2} 
& $\tfrac{2}{n-1}$
& $1<n$
& $1<n<\tfrac{5}{3}$
& $1<n<3$
& $1<n<\tfrac{3}{2}$
\\
\hline
sn 
& \eqref{sn1}
& $\tfrac{2}{1-n}$
& $\tfrac{1}{2}<n<1$
& $\tfrac{1}{2}<n<1$
& $\tfrac{1}{2}<n<1$
& $\tfrac{1}{2}<n<1$
\\
& \eqref{sn2} 
& $\tfrac{2}{n-1}$
& $1<n$
& $1<n<\tfrac{5}{3}$
& $1<n<3$
& $1<n<\tfrac{3}{2}$
\\
\hline
\hline
rational cn
& \eqref{ratcn1}, \eqref{ratcn2}, \eqref{ratcn6}
& $\tfrac{2}{n-1}$
& $1<n$
& $1<n<\tfrac{5}{3}$
& $1<n$
& $1<n<\tfrac{3}{2}$ 
\\
& \eqref{ratcn3}
& $\tfrac{1}{1-n}$
& $\tfrac{2}{3}<n<1$
& $\tfrac{2}{3}<n<1$
& $\tfrac{2}{3}<n<1$
& $\tfrac{3}{4}<n<1$
\\
& \eqref{ratcn4}, \eqref{ratcn5}
& $\tfrac{4}{1-n}$
& $\tfrac{1}{3}<n<1$
& $\tfrac{1}{3}<n<1$
& $\tfrac{1}{3}<n<1$
& $\tfrac{1}{3}<n<1$
\\
\hline
\end{tabular}
}
\caption{Existence conditions of the symmetric compactons}
\label{table:conditions}
\end{table}

\subsection{Examples when compactons exist only as weak solutions}

The preceding results show that there are specific $K(m,n)$ and $\KP(m,n)$ equations
such that the only explicit compactons of the form \eqref{types} are weak solutions. 

For example: 
algebraic type \eqref{zsq1}, cosine type \eqref{cos1}, 
Jacobi cn and sn types \eqref{cn2} and \eqref{sn2}, 
and rational Jacobi cn type \eqref{ratcn6}
are admitted as weak solutions by both equations in the case 
$n=2$, $\sgn(a)=\sgn(b)$,
with $m=3/2,2,3,3,5/2$ respectively,
but fail to exist as strong solutions. 
Moreover, in each of these cases, 
all of the other types \eqref{zsq2}, \eqref{cos2}, \eqref{cn1} and \eqref{sn1}, \eqref{ratcn1}--\eqref{ratcn5} 
are not solutions since they have either $\sgn(a)=-\sgn(b)$ 
or an opposite relation between $\sgn(a)$ and $\sgn(g)$. 

Another example is the second cosine type \eqref{cos2}, which has $n=1$:
in the case $m=1/4$, it is admitted only as a weak solution by both equations,
with $\sgn(a)=-\sgn(b)$.

\subsection{Examples of numerical weak compactons}\label{sec:numerical}

The existence conditions \eqref{A1A2A3=0} and \eqref{A4=0} for weak compactons
enable finding numerical solutions of the respective 
integral equations \eqref{weak.Kmn.pde} and \eqref{weak.KPmn.pde}. 
It is most convenient to work in terms of the variable $V(\xi)= U(\xi)^n$,
which satisfies 
\begin{equation}\label{2ndordODE.V}
V'' = 2\big( C + (1+1/n)B V^{1/n} - (1+m/n)A V^{m/n} \big)
\end{equation}
obtained from differentiation of the first-order ODE \eqref{reducedODE.V}. 

In the case of the $K(m,n)$ equation, 
existence condition \eqref{A1A2A3=0} is equivalent to the cutoff conditions
\begin{equation}\label{V.V'.V''.conds}
V(\pm L)=0,
\quad
V'(\pm L)=0,
\quad
V''(\pm L)=0
\end{equation}
on solutions $V(\xi)$ of ODE \eqref{2ndordODE.V}
in an interval $-L \leq\xi\leq L$.
These conditions together require that $C=0$ and hence $E=0$, 
as seen from the ODEs \eqref{2ndordODE.V} and \eqref{reducedODE.V}, 
whereby the first condition then implies that the second and third conditions hold automatically. 
To obtain a weak compacton solution that has a symmetric profile, 
it is sufficient to take an initial condition $V(0)=V_0\neq 0$, $V'(0)=0$,
along with $\sgn(V''(0)) = - \sgn(V_0)$,
in the ODE \eqref{2ndordODE.V}. 
The initial value $V_0$ here is determined by the first-order ODE \eqref{reducedODE.V},
which yields the algebraic relation $B V_0^{1/n} -A V_0^{m/n} =0$. 
Thus, 
\begin{equation}\label{V0}
V_0 = (B/A)^{n/(m-1)} = \big(\tfrac{(n+m)g}{(n+1)a}\big)^{n/(m-1)}
\end{equation}
from expressions \eqref{Kmn.rels} for $B$ and $A$. 
The sign condition thereby becomes
\begin{equation}\label{sgnV''}
(1-m) (B^{m-n}/A^{1-n})^{1/(m-1)} = \tfrac{1-m}{b} \big(\tfrac{g^{m-n}}{a^{1-n}}\big)^{1/(m-1)} <0 . 
\end{equation}
In addition, the endpoints $\pm L$ are determined by the integral 
\begin{equation}\label{L}
L = \int_0^{V_0} \frac{dV}{\sqrt{B (1 - (V/V_0)^{(m-1)/n})V^{1+1/n}}} 
= \int_0^{V_0} \frac{dV}{\sqrt{A ((V_0/V)^{m/n}-1)V^{1+m/n}}} 
\end{equation}
through integration of ODE \eqref{reducedODE.V}.
The resulting solution of the ODE satisfies the integral equation \eqref{weak.Kmn.pde}. 

The case of the $\KP(m,n)$ equation is similar,
with the additional existence condition \eqref{A4=0} 
which is equivalent to the cutoff condition $V'''(\pm L)=0$. 
This condition holds as a consequence of the previous conditions \eqref{V.V'.V''.conds},
as seen from differentiation of the second order ODE \eqref{2ndordODE.V}. 
Thus, a symmetric weak compaction solution is given by 
solving the ODE \eqref{2ndordODE.V} with the same initial condition 
$V(0)=V_0\neq 0$, $V'(0)=0$, using relations \eqref{V0}, \eqref{sgnV''}, \eqref{L}.
The resulting solution can be straightforwardly shown to satisfy the integral equation \eqref{weak.KPmn.pde}. 

\begin{figure}[h]
\centering
\includegraphics[trim=2cm 17cm 10cm 2cm,clip,width=0.49\textwidth]{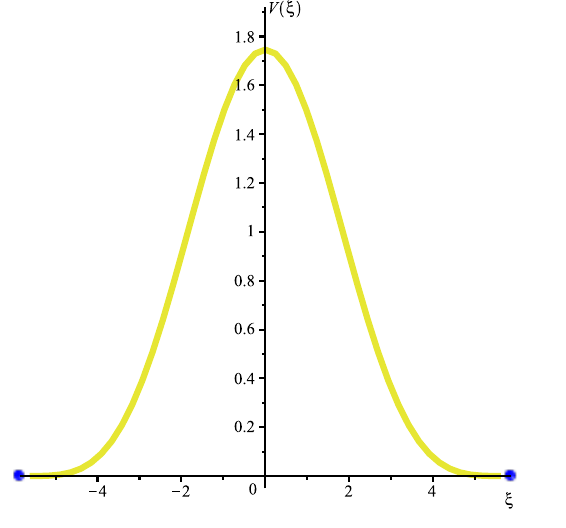}
\includegraphics[trim=2cm 17cm 10cm 2cm,clip,width=0.49\textwidth]{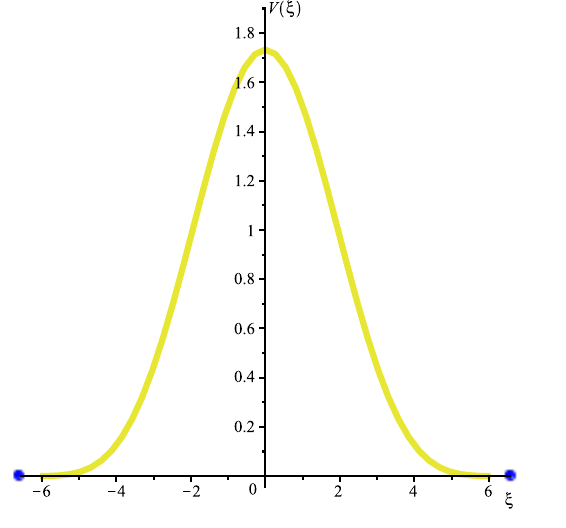}
\caption{Profile of symmetric weak compactons:
(left) $n=2$, $m=9/4$, $a=b=g=1$; 
(right) $n=9/10$, $m=1/2$, $a=-b=g=1$
}\label{fig-numerical}
\end{figure}

It is straightforward to obtain and plot the resulting numerical weak compactons
expressed in terms of the original variable $U(\xi) = V(\xi)^{1/n}$. 
Two examples are shown in \Figref{fig-numerical},
which have $m=\tfrac{1}{4}(5n-1)$ and $m=5n-4$, respectively.
(These relations have been chosen to differ from all of the ones arising
in the explicit solutions).
Numerical solution of the ODE has been done using Maple. 
For the values of $n$ used for these numerical solutions, 
explicit symmetric compacton expressions cannot be found
(i.e.\ for equations $K(9/4,2)$ and $K(1/2,9/10)$,
as well as equations $\KP(9/4,2)$ and $\KP(1/2,9/10)$).

A similar analysis can be applied to obtain numerical weak compactons 
having an antisymmetric profile.

\section{Concluding remarks}\label{sec:conclude}

Weak solutions provide a general setting for formulating and understanding compactons,
such that classical (strong) solutions are a special case. 
In this setting, there exist weak compactons that have insufficient differentiability 
to be classical solutions. 
Additionally, it enables finding numerical compacton solutions. 

New types of explicit compacton profiles have been derived 
for both the $K(m,n)$ equation \eqref{Kmn}, 
which is a generalization of the gKdV equation having nonlinear dispersion, 
and for its two-dimensional counterpart called the $\KP(m,n)$ equation \eqref{KPmn},
which is an analogous generalization of the gKP equation. 
These new compactons comprise weak solutions as well as classical solutions. 
In particular, examples of compactons that exist only as weak solutions are given.
They may be helpful in understanding the question of well-posedness for these equations
(see the recent work \cite{AkhAmbWri,GerHar-GriMar}). 

The same method is applicable more generally to other nonlinearly dispersive equations,
such as generalizations of the KdV and KP equations 
with different forms of nonlinearities \cite{RosOro,Yu}.

\section*{Acknowledgments}
SCA is supported by an NSERC Discovery grant.
MLG gratefully acknowledges support of  Junta de Andalucia FQM-201 group.
Referees are thanked for helpful remarks which have improved this paper.

\end{document}